\documentclass{article}

\usepackage{arxiv}

\usepackage{amsmath,amsfonts,amssymb}
\usepackage{array}
\usepackage{booktabs}
\usepackage{multirow}
\usepackage{tabularx}
\usepackage{graphicx}
\usepackage[caption=false,font=normalsize,labelfont=sf,textfont=sf]{subfig}
\usepackage{textcomp}
\usepackage{stfloats}
\usepackage{url}
\usepackage{verbatim}
\usepackage{cite}
\usepackage{algorithm}
\usepackage{orcidlink}
\usepackage{algpseudocode}
\usepackage{pgfplots}
\usepackage{graphicx}
\usepackage{wrapfig}
\usepackage{pgfplotstable}
\usepackage{tikz}
\usepgfplotslibrary{groupplots}
\usetikzlibrary{positioning,arrows.meta,shapes.geometric,fit}
\pgfplotsset{compat=1.18}

\title{\textbf{PrivFedTalk}: Privacy-Aware Federated Diffusion with Identity-Stable Adapters for Personalized Talking-Head Generation}

\author{
\textbf{Soumya Mazumdar}~\orcidlink{0009-0006-3521-9557}\\
Department of Computer Science and Business Systems\\
Gargi Memorial Institute of Technology\\
Affiliated to Maulana Abul Kalam Azad University of Technology\\
Balarampur, Mouza Beralia, Baruipur, Kolkata 700144, West Bengal, India\\
\texttt{reachme@soumyamazumdar.com}
\and
\textbf{Vineet Kumar Rakesh}~\orcidlink{0009-0000-7102-6564}\\
Engineering Sciences, Homi Bhabha National Institute\\
Anushaktinagar, Mumbai, Maharashtra 400094, India\\
Computer and Informatics Group / Radioactive Ion Beam Facilities Group\\
Variable Energy Cyclotron Centre\\
1/AF, Bidhannagar, Kolkata 700064, West Bengal, India\\
\texttt{vineet@vecc.gov.in}
\and
\textbf{Tapas Samanta}~\orcidlink{0000-0003-0521-0747}\\
Engineering Sciences, Homi Bhabha National Institute\\
Anushaktinagar, Mumbai, Maharashtra 400094, India\\
Computer and Informatics Group / Radioactive Ion Beam Facilities Group\\
Variable Energy Cyclotron Centre\\
1/AF, Bidhannagar, Kolkata 700064, West Bengal, India\\
\texttt{tsamanta@vecc.gov.in}
}

\begin{document}
\maketitle

\begin{abstract}
Talking-head generation has advanced rapidly with diffusion-based generative models, but training such systems typically requires centralized collections of face videos and speech, which raises serious privacy concerns. This challenge is especially severe for personalized talking-head generation, where identity-specific data are highly sensitive and often cannot be pooled across users or devices. This paper presents \textit{PrivFedTalk}, a privacy-aware federated framework for personalized talking-head generation that combines conditional latent diffusion with parameter-efficient identity adaptation. A shared diffusion backbone is coordinated across clients, while each client learns lightweight LoRA-style identity adapters from local private audio-visual data, thereby avoiding raw data sharing and reducing communication overhead. To improve performance under heterogeneous client distributions, \textit{Identity-Stable Federated Aggregation} (ISFA) weights client updates using privacy-safe scalar reliability signals derived from on-device identity consistency and temporal stability estimates. A \textit{Temporal-Denoising Consistency} (TDC) regularization strategy is introduced to constrain inter-frame drift during denoising and to reduce flicker and identity drift in federated training. To reduce update-side privacy risk, secure aggregation and client-level differential privacy are applied to adapter updates. A practical implementation supports both GPU customized-memory execution and multi-GPU client-parallel training, enabling deployment on heterogeneous shared hardware. The implementation supports stable federated optimization and successful execution of the training and evaluation pipeline under low-memory settings. A comprehensive comparative study across diverse training and aggregation conditions on the present setup was performed for PrivFedTalk, FedAvg, and FedProx, with performance assessed using the reported quantitative metrics. These results support the effectiveness and feasibility of privacy-aware personalized talking-head training in constrained federated environments, while indicating that stronger component-wise, privacy–utility, and qualitative claims require additional standardized evaluation. GitHub: \url{https://github.com/mazumdarsoumya/PrivFedTalk}
\end{abstract}


\section{Introduction}

Diffusion models have emerged as high-fidelity generative mechanisms for image synthesis by learning to invert a gradual noise corruption process \cite{ho2020ddpm}. Latent diffusion further reduces computational cost by moving the diffusion process to a learned latent space \cite{rombach2022ldm}. In parallel, talking-head generation methods have advanced significantly in speech-driven facial animation and lip synchronization \cite{zhou2020makeittalk,prajwal2020wav2lip,ren2023hrnet,lyu2025etau}. However, many training pipelines require centralized identity-bearing data, which is often infeasible in privacy-sensitive settings.

Federated learning offers a mechanism for collaborative training without centralizing raw data~\cite{chen2025pfedlah,an2025privacyretrieval}. The canonical baseline is iterative model averaging (FedAvg) \cite{mcmahan2017fedavg}, while heterogeneity-aware extensions such as FedProx address non-Independent and Identically Distributed (non-IID) and system variability \cite{li2020fedprox}. Update-side leakage risk can be reduced through secure aggregation \cite{bonawitz2017secureagg} and differential privacy \cite{dwork2006calibrating,geyer2017clientdp,mcmahan2018dprnn}. The remaining challenge is to combine these components in a form suitable for temporally stable, identity-preserving talking-head generation under strong client heterogeneity.

\subsection{Background and Motivation}
Personalized talking-head generation is attractive for privacy-aware avatars, telepresence, assistive communication, and identity-preserving human--computer interaction. However, identity-bearing face videos and speech are sensitive biometric signals. Centralized training therefore creates a privacy bottleneck. A federated formulation is a natural alternative, but standard federated optimization is not sufficient by itself because talking-head generation requires identity preservation, lip synchronization, perceptual quality, and temporal stability at the same time.

\subsection{Challenges in Federated Talking-Head Generation}
Four major challenges arise in federated personalized talking-head generation.

First, client data are naturally non-IID. Each client may contain only a narrow subset of identities and restricted local variation. Uniform aggregation therefore becomes fragile because locally useful updates may become globally harmful when averaged without quality awareness.

Second, temporal coherence matters as much as frame realism. A talking-head sequence must preserve smooth lip motion, stable pose progression, and persistent identity cues across time. Minor framewise inconsistencies in local updates can accumulate into noticeable temporal artifacts.

Third, privacy risk remains present even if raw data do not leave the client. Model updates themselves may leak information unless additional protection is used.

Fourth, diffusion models are large. Full-model transmission is expensive and often impractical in federated settings. A communication-efficient alternative is therefore necessary.

\subsection{Problem Statement}
The goal is to train a personalized talking-head generator under federated privacy constraints. A central server coordinates optimization across distributed clients, but each client retains raw audio-visual data locally. The model must synthesize identity-consistent and temporally stable talking-head videos conditioned on speech-related inputs, without centralizing raw face or voice data.

The central question is therefore the following: how can a diffusion-based personalized talking-head generator be trained in a federated manner while preserving privacy, tolerating non-IID client distributions, maintaining temporal stability, and limiting communication cost?

\subsection{Overview of PrivFedTalk}
PrivFedTalk addresses this problem through four main design choices.

First, a shared conditional diffusion backbone captures general speech-to-face generation structure. Second, compact client-local Low-Rank Adaptation (LoRA)-style identity adapters encode personalization while keeping communication small. Third, a Temporal-Denoising Consistency regularizer constrains inter-frame variation during denoising. Fourth, Identity-Stable Federated Aggregation weights client contributions using privacy-safe quality signals instead of uniform averaging.

A practical privacy stack is also included through secure aggregation and client-level differential privacy over adapter updates.

\subsection{Main Contributions}
The main contributions are summarized as follows:
\begin{enumerate}
    \item A privacy-aware federated framework for personalized talking-head generation based on a shared conditional diffusion backbone and client-local low-rank identity adapters.
    \item A Temporal-Denoising Consistency regularizer that constrains frame-to-frame drift during denoising and is designed to improve temporal stability.
    \item Identity-Stable Federated Aggregation, which weights client updates using privacy-safe scalar signals derived from local identity consistency and temporal stability.
    \item A practical PyTorch implementation that supports both GPU customized-memory execution and multi-GPU client-parallel execution, with configurable device selection and memory-aware settings.
\end{enumerate}

\section{Related Work}

\subsection{Talking-Head Generation and Lip Synchronization}
Audio-driven talking-head generation has been studied across classical reenactment, portrait animation, neural rendering, and direct synthesis settings. Early face reenactment and portrait animation systems include graphics- and GAN-based approaches such as Face2Face, First Order Motion Model, and MakeItTalk \cite{thies2016face2face,zhou2020firstorder,zhou2020makeittalk}. Subsequent works improved controllability and fidelity through semantic neural rendering and 3D-aware representations~\cite{ren2023hrnet,lyu2025etau}, including AD-NeRF, PIRenderer, deformable NeRF-based formulations, and more recent Gaussian-based avatar and talking-head models \cite{guo2021adnerf,ren2021pirenderer,li2023deformablenerf,ma2024gaussianblendshapes,gaussiantalker2024}. Lip synchronization remains a central requirement in speech-driven portrait animation. Wav2Lip proposes a lip synchronization expert that improves speech-to-lip generation under unconstrained conditions \cite{prajwal2020wav2lip}, while SyncNet-style audio-visual synchronization metrics are grounded by Chung and Zisserman \cite{chung2016outoftime}. Diffusion-based talking-face priors increasingly improve performance and facial texture realism in challenging speech-driven settings \cite{ani2022facetalk,iccv2023audiovisualdiffusion,animateme2024}. For a broader taxonomy of talking-head generation paradigms, datasets, and open challenges, see the recent survey in \cite{rakesh2025thgreview,chen2025avsac}.

\subsection{Diffusion and Latent Diffusion}
Denoising diffusion probabilistic models formalize the forward noising process and reverse denoising learning objective for high-quality generation \cite{ho2020ddpm}. Latent diffusion models improve efficiency by carrying out the diffusion process in a learned latent manifold while preserving synthesis quality \cite{rombach2022ldm}. These formulations are especially attractive for talking-head generation because they offer strong generative fidelity while remaining more scalable than full-pixel diffusion for video-related synthesis tasks.

\subsection{Federated Learning and Privacy}
Federated learning provides a framework for decentralized optimization without centralizing raw client data. FedAvg is the standard baseline in federated optimization \cite{mcmahan2017fedavg}, while FedProx and SCAFFOLD address client heterogeneity and update drift through proximal regularization and control variates, respectively \cite{li2020fedprox,karimireddy2020scaffold}. Secure aggregation prevents the server from observing individual client updates and reveals only protected aggregates \cite{bonawitz2017secureagg}. Differential privacy bounds information leakage through sensitivity-aware noise injection and is foundational both in centralized and federated settings \cite{abadi2016dpsgd,dwork2014dp,geyer2017clientdp,mcmahan2018dprnn}. At the same time, federated optimization remains vulnerable to inference and poisoning attacks, including membership inference, gradient leakage, and backdoor manipulation \cite{shokri2017membership,zhu2019dlg,bagdasaryan2020backdoor}. These risks motivate privacy-aware and performance-aware aggregation strategies for sensitive generative applications.

\subsection{Parameter-Efficient Adaptation for Personalized Generation}
Parameter-efficient adaptation methods reduce memory cost, communication overhead, and optimization instability by updating only a compact subset of effective parameters. Adapter-based tuning and low-rank adaptation have been widely studied as efficient alternatives to full fine-tuning \cite{rebuffi2018adapters,he2021towards,hu2022lora,jian2024unifrd}. This family of methods is especially attractive in federated personalized generation because the shared backbone can preserve common generative knowledge while client-specific adapters capture identity-dependent information locally~\cite{chen2025pfedlah,an2025privacyretrieval}. In this work, we leverage this principle for communication-efficient and privacy-aware identity personalization in talking-head generation.

\section{Problem Setting and Threat Model}

\subsection{Federated Personalized Talking-Head Learning Setting}
Consider a federated talking-head generation system with $K$ clients, where each client corresponds to a user or device that stores private audio-visual data locally. Client $k$ holds a dataset
\begin{equation}
\mathcal{D}_k=\{(x_k^i,a_k^i)\}_{i=1}^{n_k},
\end{equation}
where $x_k^i$ denotes a face video clip and $a_k^i$ denotes the associated conditioning signal, such as audio features or phoneme-related information. The full training data are distributed across clients and are not centralized.

A central server coordinates federated optimization over communication rounds $t=1,\dots,T$. In each round, only a subset of clients participates, reflecting realistic partial participation and client dropout. Raw videos, raw speech, and identity-sensitive inputs remain on-device throughout training. The server only receives privacy-protected adapter updates together with small scalar signals required for weighted aggregation.

The task is personalized talking-head generation under privacy constraints. The model must produce a realistic output face video that follows the conditioning speech content while preserving the target identity and maintaining temporal smoothness over time. This is difficult because client data are naturally non-IID across identity, pose, expression, and speaking style.

\subsection{Threat Model}
The server is assumed to be honest-but-curious. It follows the training protocol correctly but may attempt to infer sensitive information about client data from communicated updates. This is especially important in talking-head generation because faces, voices, and identity-specific motion cues are sensitive biometric signals.

The server does not access raw face videos, raw audio, or reference identity images. However, privacy leakage may still occur if model updates reveal information about local data. The method therefore treats client updates as potentially sensitive objects and protects them using secure aggregation and client-level differential privacy.

Client dropout is also part of the threat and deployment model. In practical federated systems, only a fraction of clients may participate in each round due to availability, connectivity, or resource limits. performance tests may additionally include unreliable or adversarial clients.

\subsection{System Design Objectives}
Based on this setting, the method is designed around four goals:
\begin{enumerate}
    \item \textbf{Privacy preservation:} training should avoid centralizing raw face and voice data while limiting leakage through communicated updates.
    \item \textbf{Identity preservation:} the generated talking head should remain visually consistent with the target identity.
    \item \textbf{Temporal stability:} long generated videos should avoid flicker, frame-to-frame jitter, and identity drift.
    \item \textbf{Federated performance and efficiency:} the system should remain effective under non-IID client distributions, partial participation, and constrained communication.
\end{enumerate}

\section{PrivFedTalk Method}

\subsection{Overview of the Framework}

PrivFedTalk combines a shared conditional diffusion backbone with client-local low-rank identity adapters. The backbone captures general audio-to-face generation dynamics, while the adapters absorb user-specific identity information in a communication-efficient way. This design avoids full-model transmission and keeps personalization close to the client.

During each communication round, the server broadcasts the current global model state. Each selected client updates only its local adapter parameters using private audio-visual data and the multi-term objective in \eqref{eq:full_objective}. The client then computes a clipped and optionally noised adapter update, sends it through secure aggregation, and also reports a privacy-safe scalar reliability score derived from local identity and temporal quality checks. The server aggregates the protected client updates using identity-stable weighting.

Let $\theta$ denote the shared conditional diffusion backbone and let $\phi_k$ denote the client-local low-rank identity adapter for client $k$. The global adapter state at communication round $t$ is denoted by $\phi^t$. For each training example, $x$ denotes a face video clip and $a$ denotes the corresponding speech-driven conditioning input. The embedding extracted from audio or phoneme information is written as $c(a)$. Let $r$ denote an identity reference image and let $e(r)$ denote the identity embedding extracted from $r$. During federated optimization, client $k$ sends an adapter update $\Delta \phi_k$, and its clipped and noised version is denoted by $\tilde{\Delta}\phi_k$. The server may further use a client reliability score $s_k$ to compute an aggregation weight $w_k$. We denote by $p$ the client participation fraction in each communication round and by $E$ the number of local training epochs.

\begin{figure}
    \centering
    \includegraphics[width=1\linewidth]{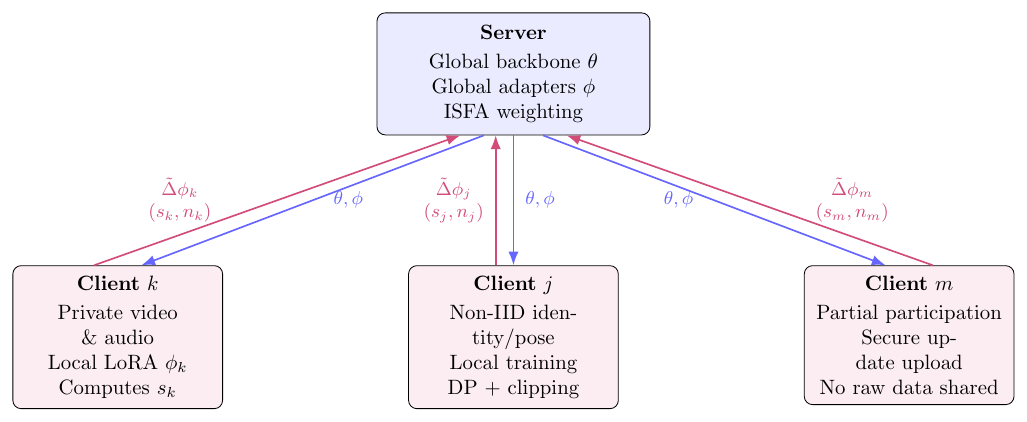}
    \caption{System-level overview of PrivFedTalk. The server maintains the shared diffusion backbone $\theta$ and the current global adapter state $\phi$, while each participating client performs private local adaptation using lightweight LoRA modules. Only protected adapter updates $\tilde{\Delta}\phi$ and privacy-safe scalar reliability signals $(s,n)$ are returned to the server, which applies identity-stable federated aggregation (ISFA) to produce the next global adapter.}
    \label{fig:framework}
\end{figure}

Figure~\ref{fig:framework} summarizes the system-level workflow of PrivFedTalk. The server maintains a shared diffusion backbone and the current global adapter state, while each participating client performs local personalization using lightweight LoRA modules over private audio-visual data. Rather than transmitting raw face or speech samples, each client returns only a protected adapter update together with a privacy-safe scalar reliability signal derived from identity preservation and temporal stability. The server then combines these protected updates through identity-stable federated aggregation, so that clients producing more reliable local behavior contribute more strongly to the next global adapter.

\subsection{Conditional Latent Diffusion Backbone}
In the latent diffusion space, the clean latent video is denoted by $z_0$, while $z_t$ denotes the corresponding noisy latent at diffusion step $t$. The diffusion variance schedule is written as $\beta_t$, with $\alpha_t=1-\beta_t$ and cumulative coefficient $\bar{\alpha}_t=\prod_{s=1}^{t}\alpha_s$. The forward diffusion process is defined by
\begin{equation}
q(z_t \mid z_{t-1})=\mathcal{N}\!\left(\sqrt{1-\beta_t}\,z_{t-1},\beta_t I\right),
\label{eq:forward_step}
\end{equation}
and equivalently
\begin{equation}
q(z_t \mid z_0)=\mathcal{N}\!\left(\sqrt{\bar{\alpha}_t}\,z_0,(1-\bar{\alpha}_t)I\right),
\label{eq:forward_closed}
\end{equation}
where
\begin{equation}
\alpha_t=1-\beta_t,\qquad \bar{\alpha}_t=\prod_{s=1}^{t}\alpha_s.
\end{equation}

The denoiser predicts the injected Gaussian noise:
\begin{equation}
\epsilon_{\theta}(z_t,t,c(a),e(r))\approx \epsilon.
\label{eq:denoiser}
\end{equation}

The standard diffusion training objective is
\begin{equation}
\mathcal{L}_{\mathrm{diff}}(\theta)=
\mathbb{E}_{z_0,\epsilon,t}\Big[\|\epsilon-\epsilon_{\theta}(z_t,t,c,e)\|_2^2\Big].
\label{eq:ldiff}
\end{equation}
The latent diffusion setting follows the standard motivation for efficiency \cite{rombach2022ldm}. The denoiser predicts the injected noise as in DDPM-style training \cite{ho2020ddpm}.

\subsection{Client-Local LoRA Identity Adapters}
Instead of federating the full backbone, client-local low-rank adapters are trained. For a weight matrix $W$, an adapted weight is
\begin{equation}
W' = W + \Delta W,\qquad \Delta W = BA,
\end{equation}
where $A\in\mathbb{R}^{r\times d}$ and $B\in\mathbb{R}^{d\times r}$ with rank $r\ll d$ \cite{hu2022lora}. This reduces communication and isolates identity-specific personalization to compact modules.

\subsection{Temporal-Denoising Consistency}
Let $\epsilon^{(f)}$ denote the injected noise for frame $f$ and let
\[
\Delta \epsilon^{(f)}=\epsilon^{(f)}-\epsilon^{(f-1)}.
\]
Temporal-Denoising Consistency penalizes mismatch between temporal differences in true and predicted noise:
\begin{equation}
\mathcal{L}_{\mathrm{tdc}}=
\mathbb{E}\left[\|\Delta \epsilon^{(f)}-\Delta \epsilon_{\theta}(z_t,t,c,e)^{(f)}\|_1\right].
\end{equation}

\subsection{Perceptual, Identity, and Lip-Sync Losses}
The diffusion loss alone is not sufficient for high-quality personalized talking-head generation. Additional frozen networks are used locally to enforce identity consistency, perceptual similarity, and speech alignment.

The identity loss is defined as
\begin{equation}
\mathcal{L}_{\mathrm{id}}=
1-\cos\!\big(g(\hat{x}),g(x_{\mathrm{ref}})\big),
\end{equation}
where $g(\cdot)$ is a face embedding network and $\hat{x}$ denotes the generated video frame or clip.

The perceptual loss is
\begin{equation}
\mathcal{L}_{\mathrm{perc}}=
\|h(\hat{x})-h(x)\|_1,
\end{equation}
where $h(\cdot)$ is a perceptual feature extractor.

The lip-sync loss is written as
\begin{equation}
\mathcal{L}_{\mathrm{sync}}=
\mathrm{SyncLoss}(\hat{x},a),
\end{equation}
where the sync model measures audio-visual alignment.

At the client level, the local optimization target combines diffusion reconstruction, temporal regularization, identity preservation, perceptual similarity, and lip-sync supervision:
\begin{equation}
\mathcal{L}_k=
\mathcal{L}_{\mathrm{diff}}
+\lambda_{\mathrm{tdc}}\mathcal{L}_{\mathrm{tdc}}
+\lambda_{\mathrm{id}}\mathcal{L}_{\mathrm{id}}
+\lambda_{\mathrm{perc}}\mathcal{L}_{\mathrm{perc}}
+\lambda_{\mathrm{sync}}\mathcal{L}_{\mathrm{sync}}.
\label{eq:full_objective}
\end{equation}

Accordingly, the federated learning objective can be written as
\begin{equation}
\min_{\theta,\{\phi_k\}}
\sum_{k=1}^{K}\frac{n_k}{\sum_{j=1}^{K}n_j}\,
F_k(\theta,\phi_k),
\qquad
F_k(\theta,\phi_k)=
\mathbb{E}_{(x,a)\sim\mathcal{D}_k}[\mathcal{L}_k],
\end{equation}
subject to the constraint that raw client data never leave local devices.

\subsection{Privacy Stack and Identity-Stable Federated Aggregation}

Client updates are clipped and noised to provide client-level differential privacy:
\[
\tilde{\Delta \phi_k} =
\frac{\Delta \phi_k}{\max \left(1,\frac{\|\Delta \phi_k\|_2}{C}\right)}
+ \mathcal{N}(0,\sigma^2 C^2 I),
\tag{19}
\]
where \(C\) is the clipping norm and \(\sigma\) is the noise multiplier. Secure aggregation is used so that the server observes only protected aggregate adapter updates rather than individual client deltas.

Each participating client also computes a privacy-safe scalar reliability signal
\[
s_k = \alpha \, IDSim_k + (1-\alpha)\, TempStab_k,
\tag{20}
\]
where \(IDSim_k \in [0,1]\) denotes local identity consistency measured on held-out client samples using a frozen face-embedding model, and \(TempStab_k \in [0,1]\) denotes a normalized local temporal-stability statistic computed from generated validation clips. Only the scalar pair \((s_k,n_k)\) is transmitted, where \(n_k\) is the local sample count used for weighting.

The server aggregates protected adapter updates using
\[
w_k =
\frac{n_k \exp(\gamma s_k)}
{\sum_{j \in S_t} n_j \exp(\gamma s_j)},
\qquad
\phi^{t+1} = \phi^t + \eta \sum_{k \in S_t} w_k \tilde{\Delta \phi_k},
\tag{21}
\]
where \(\gamma\) controls the sharpness of quality-aware weighting and \(\eta\) is the server update scale. This formulation preserves the usual data-size dependence of federated averaging while giving larger influence to client updates that are simultaneously more identity-consistent and temporally stable under local validation.

\subsection{Federated Training Procedure}

The full round-wise optimization protocol is summarized in Algorithm~\ref{alg:privfedtalk}. In each communication round, the server samples a subset of clients, broadcasts the shared backbone and current global adapter, and lets the selected clients optimize only their local LoRA parameters. Each client then clips and optionally perturbs its adapter update before secure upload, together with the privacy-safe scalar pair needed for weighted aggregation. This procedure makes the method practical for heterogeneous federated deployment because the communication object is small, the personalization step remains local, and the global server update remains compatible with secure aggregation.

\begin{algorithm}[t]
\caption{PrivFedTalk Federated Training}
\label{alg:privfedtalk}
\begin{algorithmic}[1]
\Require Backbone $\theta$, global adapter $\phi^0$, rounds $T$, client fraction $p$
\For{$t=0$ to $T-1$}
    \State Sample participating clients $S_t$
    \ForAll{$k\in S_t$ \textbf{in parallel}}
        \State Receive $(\theta,\phi^t)$ and initialize $\phi_k\leftarrow\phi^t$
        \State Optimize $\phi_k$ locally using $\mathcal{L}_{\mathrm{diff}}+\lambda_{\mathrm{tdc}}\mathcal{L}_{\mathrm{tdc}}+\lambda_{\mathrm{id}}\mathcal{L}_{\mathrm{id}}+\lambda_{\mathrm{perc}}\mathcal{L}_{\mathrm{perc}}+\lambda_{\mathrm{sync}}\mathcal{L}_{\mathrm{sync}}$
        \State Compute update $\Delta\phi_k=\phi_k-\phi^t$
        \State Clip and noise $\Delta \phi_k$ to get $\tilde{\Delta \phi_k}$
        \State Compute local scalar $s_k = \alpha \, IDSim_k + (1-\alpha)\, TempStab_k$
        \State Send $\tilde{\Delta \phi_k}$ via secure aggregation and send scalar $(s_k, n_k)$
    \EndFor
    \State Server computes $w_k \propto n_k \exp(\gamma s_k)$ and aggregates to produce $\phi^{t+1}$
\EndFor
\end{algorithmic}
\end{algorithm}

Algorithm~\ref{alg:privfedtalk} also clarifies the separation between client-local optimization and server-side aggregation: the local objective is optimized only on-device, while the server receives protected adapter deltas rather than raw training samples. This distinction is important for both privacy claims and implementation clarity.

\subsection{Multi-GPU Execution}
The implementation supports both single-GPU and multi-GPU federated execution, can be pinned to selected visible devices such as \texttt{cuda:0} or \texttt{cuda:1}, and can be adapted to the currently available GPU memory budget through configurable batch size, LoRA rank, evaluation budget, dataloader worker count, and mixed precision.

\subsection{Execution Pipeline}

\begin{figure}
    \centering
    \includegraphics[width=\linewidth]{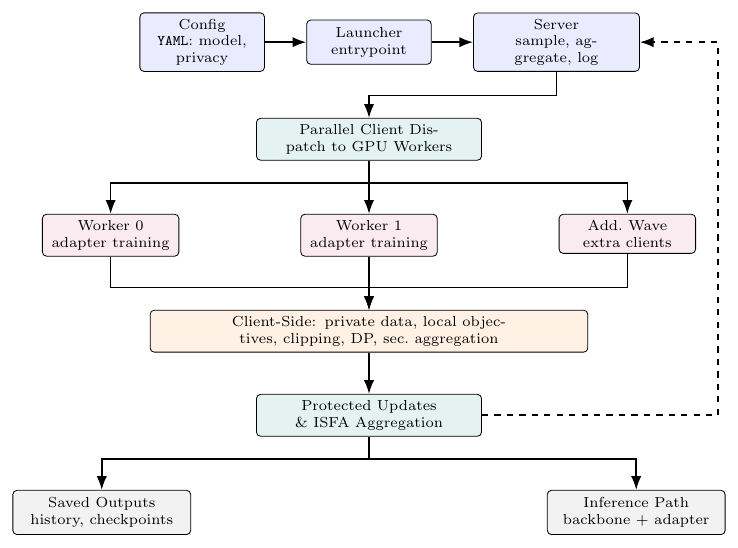}
    \caption{Practical execution pipeline of the PrivFedTalk implementation. Configuration files specify model, federation, privacy, and runtime settings. The launcher initialises the training job; the server handles client sampling, logging, and ISFA-based aggregation. Selected clients are dispatched to GPU workers for adapter-only local training, with additional clients handled in subsequent waves. Client-side processing covers private-data loading, local objective computation, gradient clipping, differential privacy, and secure aggregation. The aggregated state is fed back to the server and written to checkpoints for personalised inference via the shared backbone and a client-specific adapter.}
    \label{fig:code_pipeline}
\end{figure}

Figure~\ref{fig:code_pipeline} translates the abstract federated algorithm into the concrete execution path used in the implementation. Configuration files first define the model, federation, privacy, and runtime parameters; a launcher script instantiates the experiment; and the server process performs client sampling, aggregation, and logging. Participating clients are then mapped onto visible GPU workers, with additional clients executed in successive waves when the number of selected clients exceeds the number of available GPUs. After local processing, the protected updates return to the server for ISFA aggregation, and the resulting checkpoints are saved for later evaluation and personalized inference.

\subsection{Inferencing}
At inference time, a client uses the trained shared backbone $\theta$ together with either the final aggregated adapter or its local personalized adapter. Given a speech signal $a$ and an identity reference image $r$, the model computes the conditioning embedding $c(a)$ and the identity embedding $e(r)$.

Starting from Gaussian latent noise $z_T\sim\mathcal{N}(0,I)$, the model applies iterative reverse denoising conditioned on $c(a)$ and $e(r)$:
\begin{equation}
z_{t-1} \leftarrow \Psi_{\theta,\phi_k}(z_t,t,c(a),e(r)),
\qquad t=T,\dots,1,
\end{equation}
where $\Psi_{\theta,\phi_k}$ denotes the reverse diffusion update. The final latent $\hat{z}_0$ is decoded into the synthesized talking-head video $\hat{x}$.

\section{Experimental Setup}

\subsection{Datasets and Client Partition}
Experiments are conducted primarily on the LRS3 dataset~\cite{afouras-2018}, a large-scale audio-visual speech corpus collected from TED and TEDx videos that provides substantial diversity in speaker identity, head pose, facial appearance, illumination, and speaking style. This diversity makes LRS3 well suited for evaluating federated personalized talking-head generation under realistic cross-client variation. A federated client partition is constructed to preserve data locality and to induce non-independent and identically distributed (non-IID) heterogeneity across identities, poses, and appearance distributions. Each client retains its local audio-visual samples throughout training, and the server observes only privacy-protected adapter updates together with aggregation statistics required for federated optimization. In addition, for a limited number of runs, small subsets or selected components of HDTF~\cite{9578462} were merged with various sampled LRS3 clients to further increase client-side variability and to examine behavior under mixed audio-visual data conditions. Under the present LRS3 comparison protocol, a 2-GPU federated run was completed for 1000 communication rounds with 10 sampled clients per round.

\subsection{Preprocessing Pipeline}
Each client stores private face video and audio pairs $(x,a)$. The video clip is mapped into latent video space, the speech signal is converted into a conditioning representation $c(a)$, and an identity reference image $r$ or embedding $e(r)$ is provided to the generator.

Auxiliary local supervision is also computed on-device. Identity features are extracted through a frozen face embedding network, perceptual similarity is computed through a frozen perceptual backbone, and lip-sync quality is measured through a frozen audio-visual synchronization model. Because all of these operations are local, raw identity data do not leave the client device.

\subsection{Baselines}

The comparison set considered in this study comprises the following baseline families: (1) a centralized diffusion talking-head generator as an upper-bound reference, (2) FedAvg on adapters as a standard federated baseline, (3) FedProx on adapters as a heterogeneity-aware federated baseline, and (4) PrivFedTalk as the full method with ISFA, TDC, and the privacy stack.

The main quantitative comparison includes only those methods whose training and evaluation pipelines were fully verified under the same protocol. Under the present LRS3 comparison, the completed runs are PrivFedTalk, FedAvg on adapters, and the repository’s current 2-GPU FedProx configuration. The centralized diffusion upper bound remains an intended reference and is excluded from the main quantitative table until a matched reproduced run under the same preprocessing and evaluation protocol is available.

% \subsection{Evaluation}

% The target evaluation set includes identity similarity, lip-sync error, perceptual similarity, distributional quality, temporal stability, and communication cost. The verified validation pipeline directly reports validation diffusion loss, validation identity similarity, and validation temporal stability. The verified publication-evaluation pipeline reports center-frame PSNR, SSIM, LPIPS, FID, KID, face-embedding identity similarity, and full-video temporal jitter.

% An internal sync-proxy score is also available for implementation debugging and relative inspection. However, because this proxy is not a standard SyncNet/LSE-style benchmark, it is not used as a headline lip-sync result in the main paper. Standard lip-sync evaluation should be reported separately once the final SyncNet/LSE evaluation path is completed.

\subsection{Implementation}
The implementation is developed in PyTorch and follows the adapter-based federated training protocol described in Algorithm~\ref{alg:privfedtalk}, and the experiments are conducted on NVIDIA L40S GPUs with 48\,GB GDDR6 with ECC under Red Hat Enterprise Linux (RHEL). The shared diffusion backbone is instantiated from configuration files, while compact low-rank identity adapters are optimized locally at participating clients.
Mixed-precision execution is used in practice through bfloat16 arithmetic in order to reduce memory consumption while maintaining stable optimization.

The training system supports both single-GPU and multi-GPU execution. In the multi-GPU setting, one client worker process is mapped to one visible GPU, and selected clients are processed in parallel in waves whose width matches the number of visible GPUs.
In the single-GPU setting, the same training pipeline remains usable by setting the number of active GPU workers to one.

Device selection is controlled through the visible-device mask and the configured number of runtime GPU workers.
In practice, the implementation can be pinned to a specific visible device such as \texttt{cuda:0} or \texttt{cuda:1}.
The memory footprint is controlled through local batch size, evaluation batch size, LoRA rank, dataloader worker count, the number of evaluation batches, and mixed precision.
Therefore, the implementation can be adapted to the currently available GPU memory budget on shared hardware.

The implementation has been tested in both 1-GPU and multi-GPU customized-memory settings. Under the multi-GPU customized-memory PrivFedTalk run, the best observed checkpoint occurred at communication round 97, with validation loss 1.237751, validation identity similarity 0.7339, and validation temporal stability 0.9698.
In addition, a longer 2-GPU LRS3 comparison was completed for 1000 communication rounds with 10 sampled clients per round and one local epoch per client update, and this run is used only where explicitly referenced in the comparison tables.

\section{Experimental Results}

The target evaluation set includes identity similarity, lip-sync error, perceptual similarity, distributional quality, temporal stability, and communication cost. Within this setup, the verified validation pipeline directly reports validation diffusion loss, validation identity similarity, and validation temporal stability, while the verified publication-evaluation pipeline reports center-frame PSNR, SSIM, LPIPS, FID, KID, face-embedding identity similarity, and full-video temporal jitter. An internal sync-proxy score is also available for implementation debugging and relative inspection. However, because this proxy is not a standard SyncNet/LSE-style benchmark, it is not used as a headline lip-sync result in the main paper. Standard lip-sync evaluation should therefore be reported separately once the final SyncNet/LSE evaluation path is completed.

\begin{figure*}[t]
\centering
\begin{tikzpicture}
    \begin{groupplot}[
        group style={
            group size=3 by 1,
            horizontal sep=1.6cm,
        },
        width=0.32\textwidth,
        height=0.25\textwidth,
        xlabel={Communication Round},
        xlabel style={font=\small},
        ylabel style={font=\small, yshift=2pt},
        title style={font=\normalsize\bfseries, yshift=4pt},
        tick label style={font=\footnotesize},
        grid=major,
        grid style={dotted, gray!50},
        every axis plot/.append style={thick},
        xmin=1, xmax=100,
        xtick={1,20,40,60,80,100},
        legend style={
            font=\footnotesize,
            at={(0.98,0.95)},
            anchor=north east,
            draw=gray!40
        },
        clip=true,
    ]

    \nextgroupplot[
        title={Validation Loss $\downarrow$},
        ylabel={Loss},
    ]
    \addplot[blue!70!black] 
        table [x=round, y=val_loss, col sep=comma] {val_metrics_all_rounds.csv};

    \nextgroupplot[
        title={Identity Similarity $\uparrow$},
        ylabel={ID Similarity},
    ]
    \addplot[teal!80!black] 
        table [x=round, y=val_identity, col sep=comma] {val_metrics_all_rounds.csv};

    \nextgroupplot[
        title={Temporal Stability $\uparrow$},
        ylabel={Temporal Stability},
    ]
    \addplot[purple!70!black] 
        table [x=round, y=val_temporal, col sep=comma] {val_metrics_all_rounds.csv};

    \end{groupplot}
\end{tikzpicture}
\caption{Validation curves over 100 federated communication rounds for PrivFedTalk. The three plots show validation loss (left), identity similarity (centre), and temporal stability (right). The best checkpoint is observed at round 97, just before convergence, reflecting typical stochastic behaviour under heterogeneous client participation.}
\label{fig:val_metrics_all_rounds}
\end{figure*}
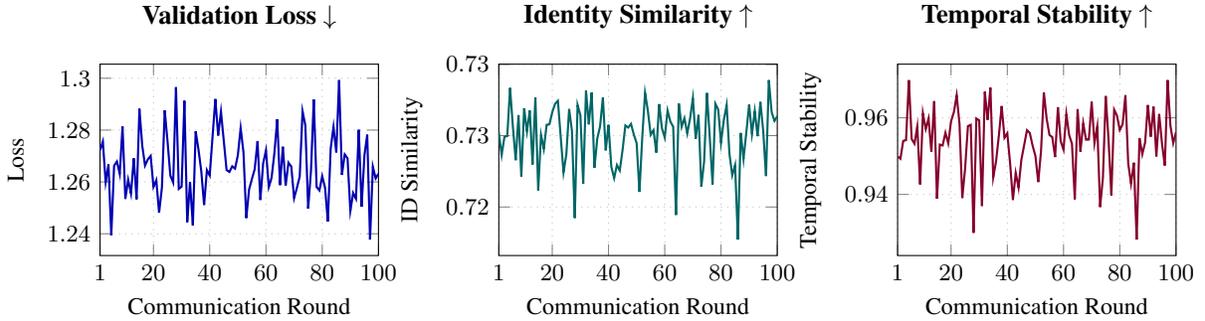

Across federated communication rounds, the validation loss, validation identity similarity, and validation temporal stability together provide an implementation-level view of optimization behavior, indicating stable training dynamics, improving identity-related validation behavior, and consistently high temporal stability as shown in Figure~\ref{fig:val_metrics_all_rounds}. The best verified checkpoint is obtained at round 97, where the validation loss reaches 1.237751, the identity similarity reaches 0.7339, and the temporal stability reaches 0.9698.

The verified training logs show stable federated optimization under the low-memory multi-GPU execution setting. The validation trajectory indicates that the strongest checkpoint is obtained before the terminal communication round, which is consistent with the stochastic nature of federated optimization under partial client participation, non-IID client sampling, and heterogeneous local update quality. Accordingly, checkpoint selection is based on the best validated model state rather than on the final communication round.

In addition to the implementation-verification run summarised in Figure~\ref{fig:val_metrics_all_rounds}, a longer 2-GPU LRS3 comparison was completed for PrivFedTalk, FedAvg on adapters, and the repository’s current 2-GPU FedProx configuration. For all three finished 1000-round runs, model selection is performed using the checkpoint with minimum validation loss. Table~\ref{tab:lrs3_bestloss_current} reports the corresponding evaluation results for those selected checkpoints.

Under the present evaluation protocol, the three methods remain numerically very close across all reported metrics. PrivFedTalk yields the lowest FID, FedAvg on adapters yields the lowest LPIPS by a marginal margin, and the current 2-GPU FedProx configuration yields the highest face-embedding identity similarity by a similarly marginal margin. These results support competitive performance under privacy-aware adapter federation, but they do not support a strong quantitative superiority claim under the current setup.

The small metric gaps in Table~\ref{tab:lrs3_bestloss_current} are also consistent with the controlled comparison setting. All three runs use the same dataset partition, the same backbone family, the same evaluation pipeline, the same checkpoint-selection rule, and closely matched low-memory training conditions. Since adaptation is restricted to a compact adapter space rather than the full backbone, the resulting optimization trajectories remain close, which limits the magnitude of separation in the final evaluation metrics.

\begin{table*}[t]
\centering
\caption{Comparison using the best validation-loss checkpoint from each finished 2-GPU 1000-round run. The FedProx row reflects the repository's current 2-GPU FedProx-config path. The internal sync-proxy metric is omitted from the main table because it is an implementation proxy rather than a standard SyncNet/LSE-style evaluation metric.}
\label{tab:lrs3_bestloss_current}
\begin{tabular}{lcccccc}
\toprule
Method & PSNR $\uparrow$ & SSIM $\uparrow$ & LPIPS $\downarrow$ & FID $\downarrow$ & ID Similarity $\uparrow$ & Temporal Jitter $\downarrow$ \\
\midrule
PrivFedTalk  & 7.3395 & 0.009683 & 1.3042 & 391.3217 & 0.018287 & 0.199548 \\
FedAvg~\cite{mcmahan2017fedavg}  & 7.3395 & 0.009683 & 1.3042 & 391.3345 & 0.018285 & 0.199548 \\
FedProx~\cite{li2020fedprox} & 7.3395 & 0.009683 & 1.3042 & 391.3342 & 0.018289 & 0.199548 \\
\bottomrule
\end{tabular}
\end{table*}

\subsection{Performance Under Non-IID Client Distributions}
The client partition protocol is explicitly non-IID by identity, appearance, pose, and speaking variation. This setting is realistic for personalized talking-head generation because each client typically contributes only a narrow and biased local distribution. Under such conditions, federated optimization can become unstable when uniformly aggregated updates are dominated by client-specific drift. The adapter-based formulation reduces this burden by restricting communication to compact personalization parameters rather than the full diffusion backbone. In addition, identity-aware weighting is intended to reduce the impact of unreliable local updates. At the present verification stage, the completed LRS3 comparison runs confirm that stable training remains feasible under heterogeneous client participation, although the currently available quantitative gaps between compared methods remain small and therefore do not yet support a strong performance superiority claim.

\subsection{Effect of Partial Client Participation}
Each communication round samples only a subset of clients, with $|S_t| \approx pK$, which matches practical federated deployment where many clients may be offline, resource-limited, or temporarily unavailable. This partial-participation regime increases stochasticity in the aggregated update and can amplify instability when local data are highly heterogeneous. The current implementation remains operational under this setting because only adapter parameters are exchanged and client updates are processed in a communication-round structure that is compatible with dynamic participation. The verified runs therefore support the practical feasibility of partial client participation, although a more detailed sensitivity analysis over participation rate should be added only after dedicated controlled experiments are completed.

\subsection{Privacy--Utility Tradeoff}

Secure aggregation and client-level differential privacy provide mechanism-level protection for communicated adapter updates, but the present study does not yet establish a complete empirical privacy account. A full privacy report should include the clipping norm, noise multiplier, client sampling rate, number of rounds, accountant assumptions, and resulting \((\varepsilon,\delta)\) guarantee, together with a privacy–utility sweep or an attack-based audit. Accordingly, the present results support privacy-aware update handling at the protocol level rather than a fully quantified privacy guarantee at the system-evaluation level.

The existing ablation structure already includes privacy-related variants, but publication-level evaluation metrics are not yet complete for every privacy ablation. Strong quantitative claims regarding the exact utility cost of privacy protection should therefore be deferred until the remaining privacy ablations and accountant-based summaries are reported under the same evaluation protocol.

\subsection{Communication Efficiency Analysis}
Communication efficiency follows directly from transmitting compact adapter updates instead of the full diffusion backbone. This reduces the amount of information exchanged in each round and makes federated personalization more practical on constrained hardware. The use of low-rank identity adapters is therefore not only a modeling choice but also a systems-level design decision that improves deployability. The verified implementation further confirms that this design is compatible with both single-GPU low-memory execution and multi-GPU client-parallel execution. A standardized communication table reporting the number of transmitted trainable parameters and the approximate bytes per communication round would strengthen reproducibility and should be included in a later revision once the corresponding parameter counts and runtime logs have been finalized under a matched protocol.

\subsection{Ablation Study}

This subsection analyses the contribution of the main design components of PrivFedTalk. The ablation protocol is organized around the current full PrivFedTalk run and a set of controlled variants with different combinations of the verified components. The currently verified variants are an adapters-only variant, a +DP variant, a +ISFA variant, a +TDC variant, and the current full run (ISFA+TDC+DP).

At the current stage, all five ablation runs provide directly comparable evaluation-side results in terms of LPIPS, FID, identity similarity, temporal jitter, and sync proxy. Table~\ref{tab:ablation_validation} reports these metrics for all currently verified ablation runs.

Table~\ref{tab:ablation_validation} shows that the +DP variant gives the lowest LPIPS and FID, while the current full run (ISFA+TDC+DP) gives the highest identity score, the lowest temporal jitter, and the highest sync proxy value. However, the margins across all five variants are very small. This indicates that the current ablation runs do not yet show strong component-wise separation on the available evaluation metrics alone.

\begin{table*}[t]
\centering
\caption{Evaluation-level ablation comparison across the currently available PrivFedTalk variants.}
\label{tab:ablation_validation}
\begin{tabular}{lccccc}
\toprule
Variant & LPIPS $\downarrow$ & FID $\downarrow$ & ID $\uparrow$ & TempJit $\downarrow$ & SyncProxy $\uparrow$ \\
\midrule
Full PrivFedTalk & 1.304183 & 391.324 & 0.01828779 & 0.199548637 & 0.083392685 \\
Adapters-only & 1.304184 & 391.322 & 0.01828694 & 0.199548656 & 0.083392585 \\
DP-only & 1.304182 & 391.320 & 0.01828744 & 0.199548647 & 0.083392678 \\
TDC-only & 1.304185 & 391.335 & 0.01828647 & 0.199548647 & 0.083392686 \\
ISFA-only & 1.304185 & 391.328 & 0.01828745 & 0.199548660 & 0.083392598 \\
\bottomrule
\end{tabular}
\end{table*}

Taken together, the current ablation results should be interpreted as preliminary component analysis rather than as definitive isolation evidence. Table~\ref{tab:ablation_validation} shows that the presently available evaluation metrics remain tightly clustered across the compared variants, so strong component-wise claims are not warranted on the basis of these results alone. The current full run is marginally strongest on identity, temporal stability, and sync proxy, while the +DP variant is marginally strongest on LPIPS and FID. Broader claims regarding the individual contributions of ISFA, TDC, client-level differential privacy, and the complete privacy stack therefore require additional fully matched ablation evidence.

\subsection{Qualitative Results}
The qualitative evaluation protocol uses matched identity-condition pairs from the test split, where each row is defined by a reference image, a target ground-truth frame, and the corresponding generated outputs from the compared models. The reference and ground-truth extraction pipeline is verified, but the final rendering path for generated baseline and PrivFedTalk outputs still requires additional validation before inclusion as a primary qualitative comparison figure. Therefore, only fully verified qualitative results should be retained in the main paper. Intermediate renderings, partial outputs, or visually noisy generations should be treated as debugging artifacts rather than as final evidence of visual quality.

\section{Discussion}

\subsection{Why Diffusion with Federated Learning}
Diffusion offers strong generative expressiveness, while federated learning offers privacy-aware distributed optimization. Their combination is particularly suitable for personalized talking-head generation because the task requires both high-capacity visual synthesis and local handling of identity-bearing data. In the present implementation, this combination was practically realizable under constrained settings, including 2-GPU execution over 1000 communication rounds with 10 sampled clients per round, while still producing stable validation behaviour. For example, the best verified checkpoint was obtained at round 97, where the validation loss reached 1.237751, validation identity similarity reached 0.7339, and validation temporal stability reached 0.9698. These results support the view that diffusion-based generation can be trained in a federated manner without destabilizing the optimization process.

\subsection{Privacy and Security Implications}
The method reduces update-side privacy risk by keeping raw videos, raw audio, and reference identity inputs on-device. Secure aggregation and client-level differential privacy further reduce the direct exposure of local adapter updates at the protocol level. This is especially relevant because the communicated object is not the full backbone but only the compact adapter update $\Delta\phi_k$, which already limits the amount of directly transmitted task-specific information. At the same time, the current study does not yet report a complete accountant-based privacy analysis with explicit \((\varepsilon,\delta)\) values, so the privacy claim remains protocol-level rather than fully quantified. The privacy-related ablation results also indicate that privacy protection does not obviously collapse utility under the present setup: for instance, the DP-only variant achieved LPIPS \(=1.304182\) and FID \(=391.320\), while the full PrivFedTalk run achieved LPIPS \(=1.304183\) and FID \(=391.324\). The small numerical gap suggests that privacy-aware update handling is feasible, but stronger privacy--utility conclusions still require a complete formal privacy report and broader controlled evaluation.

\subsection{Limitations}
The current implementation confirms stable federated optimization and practical device-aware execution, but several limitations remain. First, the final qualitative rendering path still requires full validation before generated comparison figures can be treated as publication-quality evidence. Second, the benchmark suite reported in the main paper includes only those baselines whose training and evaluation pipelines were fully verified under a matched protocol; the centralized diffusion upper bound remains an intended reference pending a matched reproduced run. Third, the completed LRS3 comparison for PrivFedTalk, FedAvg on adapters, and the current 2-GPU FedProx configuration shows only very small metric differences under the present setup and therefore supports near-parity rather than strong superiority. For example, PrivFedTalk achieved FID \(=391.3217\), FedAvg achieved FID \(=391.3345\), and FedProx achieved FID \(=391.3342\), while the identity similarity values were \(0.018287\), \(0.018285\), and \(0.018289\), respectively. Fourth, the privacy stack is implemented at the protocol level, but a complete privacy--utility characterization, including accountant-based reporting and attack-oriented validation, is still required. Finally, the reported runs use customized-memory settings for practical execution on shared hardware, so larger-scale experiments would benefit from additional efficiency, throughput, and communication-cost analysis.

\section{Conclusion}

This paper presented PrivFedTalk, a privacy-aware federated framework for personalized talking-head generation based on a shared conditional diffusion backbone, client-local low-rank identity adapters, Temporal-Denoising Consistency, and Identity-Stable Federated Aggregation. The framework was designed to address privacy risk, non-IID client heterogeneity, temporal instability, and communication efficiency in federated talking-head generation.

The present implementation demonstrates that the training pipeline operates in both single-GPU and multi-GPU customized-memory settings and can be adapted to constrained shared compute environments through configurable runtime and memory-aware settings. In the verified implementation-level run, the best checkpoint occurred at communication round 97, with validation loss \(1.237751\), validation identity similarity \(0.7339\), and validation temporal stability \(0.9698\), which supports the practical feasibility and optimization stability of the design. In addition, a longer 2-GPU LRS3 comparison was completed for 1000 communication rounds with 10 sampled clients per round, confirming that the end-to-end evaluation pipeline can be executed for PrivFedTalk and adapter-based federated baselines under the present setup.

Under this comparison protocol, the finished runs remained numerically very close. PrivFedTalk achieved PSNR \(7.3395\), SSIM \(0.009683\), LPIPS \(1.3042\), FID \(391.3217\), identity similarity \(0.018287\), and temporal jitter \(0.199548\), while the matched FedAvg and FedProx runs produced nearly identical values. The ablation study showed a similar pattern: the full PrivFedTalk variant was marginally strongest on identity score \(0.01828779\), temporal jitter \(0.199548637\), and sync proxy \(0.083392685\), whereas the DP-only variant was marginally strongest on LPIPS \(1.304182\) and FID \(391.320\). These results support competitive performance and implementation feasibility, but they do not yet establish decisive quantitative superiority or definitive component isolation.

Overall, the paper shows that privacy-aware personalized talking-head training with federated diffusion and lightweight adapters is practically achievable under constrained heterogeneous environments. At the same time, stronger claims regarding benchmark superiority, privacy--utility tradeoffs, and qualitative fidelity should be deferred to a later revision with fully matched baseline coverage, finalized lip-sync evaluation, formal privacy accounting, and broader ablation evidence.

\section*{Ethical Considerations}
Talking-head generation involves sensitive identity-bearing data and therefore requires strict attention to privacy, consent, and misuse risk. A privacy-aware training design reduces raw data exposure, but it does not eliminate the need for informed consent, controlled access, and careful downstream use. Any deployment of personalized talking-head systems should include clear data-governance policies, access control, misuse monitoring, and usage restrictions for identity-sensitive content.

\section*{Acknowledgments}
The authors gratefully acknowledge the support of the Variable Energy Cyclotron Centre (VECC), the Department of Atomic Energy (DAE), Government of India, for providing the infrastructure and technical environment that supported this research. The authors also thank the staff of the VECC library for their assistance during the course of this study.

\section*{Declarations}

\subsubsection*{Author Contributions}
Soumya Mazumdar conceived the methodology, developed the core algorithms and implementation, conducted the experiments, analysed the results, and prepared the initial manuscript draft. Vineet Kumar Rakesh contributed to the system design, technical discussion, and manuscript revision. Tapas Samanta independently verified the experimental results and analyses for technical accuracy and consistency. All authors reviewed and approved the final manuscript.

\subsubsection*{Consent to Publish}
All authors have read and approved the final manuscript and consent to its submission and publication.

\subsubsection*{Data Availability Statement}
This study uses publicly available benchmark data under the experimental protocol described in the manuscript, including the LRS3-based comparison setting. The implementation associated with this work is available at the project repository: \url{https://github.com/mazumdarsoumya/PrivFedTalk}. Additional derived experimental artifacts and implementation details are available from the corresponding author upon reasonable request.

\subsubsection*{Conflict of Interest}
The authors declare that they have no known competing financial interests or personal relationships that could have appeared to influence the work reported in this paper.

\bibliographystyle{IEEEtran}
\bibliography{privfedtalk_refs}

\section*{Author Biographies}

\renewcommand{\arraystretch}{1.2}
\noindent\begin{tabular}{@{}p{0.17\textwidth} p{0.78\textwidth}@{}}

\begin{minipage}[t]{\linewidth}
\vspace{0pt}
\includegraphics[width=\linewidth]{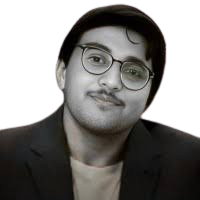}
\end{minipage}
&
\begin{minipage}[t]{\linewidth}
\vspace{0pt}
\textbf{Soumya Mazumdar} is a student researcher pursuing a B.S. in Data Science and Applications at the Indian Institute of Technology Madras and B.Tech. in Computer Science and Business Systems at West Bengal University of Technology (GMIT campus), India. His work focuses on temporal generative modeling, geometry-aware computer vision, and controllable video synthesis, with particular interest in diffusion-based methods for talking-head generation and temporal consistency. He has served as a Research Trainee at the Variable Energy Cyclotron Centre (VECC), where he worked on pose- and
landmark-conditioned video generation, benchmarking, and efficient deployment pipelines. He has contributed to research publications in journals, conference proceedings, and edited volumes, and is also associated with an Indian patent in neural network-based real-time analysis.
\end{minipage}
\\[1.5em]

\begin{minipage}[t]{\linewidth}
\vspace{0pt}
\includegraphics[width=\linewidth]{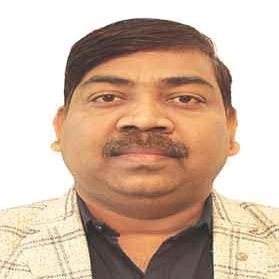}
\end{minipage}
&
\begin{minipage}[t]{\linewidth}
\vspace{0pt}
\textbf{Vineet Kumar Rakesh} is a Technical Officer (Scientific Category) at the Variable Energy Cyclotron Centre (VECC), Department of Atomic Energy, India, with over 23 years of experience in software engineering, database systems, and artificial intelligence. His research focuses on talking head generation, lip reading, and ultra-low-bitrate video compression for real-time teleconferencing. He is currently pursuing a Ph.D. at Homi Bhabha National Institute, Mumbai. Mr. Rakesh has contributed to office automation, OCR systems, and digital transformation projects at VECC. He is an Associate Member of the Institution of Engineers (India) and a recipient of the DAE Group Achievement Award.
\end{minipage}
\\[1.5em]

\begin{minipage}[t]{\linewidth}
\vspace{0pt}
\includegraphics[width=\linewidth]{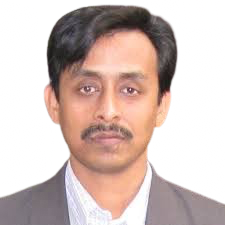}
\end{minipage}
&
\begin{minipage}[t]{\linewidth}
\vspace{0pt}
\textbf{Dr. Tapas Samanta} is a senior scientist and Head of the Computer and Informatics Group at the Variable Energy Cyclotron Centre (VECC), Department of Atomic Energy, India. With over two decades of experience, his work spans artificial intelligence, industrial automation, embedded systems, high-performance computing, and accelerator control systems. He also leads technology transfer initiatives and public scientific outreach at VECC.
\end{minipage}
\\

\end{tabular}

\end{document}